\documentclass[lettersize]{IEEEtran}
\usepackage{amsmath,amsfonts}
\IEEEoverridecommandlockouts
\normalsize

\usepackage{cite}
\usepackage{amssymb}
\usepackage{amsmath}
\usepackage{array}
\usepackage{amsthm}
\allowdisplaybreaks 
\usepackage{autobreak}
\usepackage{mathptmx}
\usepackage{mathtools, cuted}
\usepackage{algpseudocode}
\usepackage[ruled, vlined, linesnumbered]{algorithm2e}
\usepackage{graphicx}
\usepackage{subcaption} 
\usepackage{bbm}
\usepackage{booktabs}
\usepackage{multirow}
\usepackage{makecell}
\usepackage[table,xcdraw,dvipsnames]{xcolor}
\usepackage{colortbl}
\usepackage{enumitem}
\usepackage{setspace}

\usepackage[left=1.62cm,right=1.65cm,top=0.7in]{geometry}
\newtheorem{remark}{Remark}
\newtheorem{proposition}{Proposition}

\usepackage{acronym}
\acrodef{ml}[ML]{machine learning}
\acrodef{drl}[DRL]{deep reinforcement learning}
\acrodef{fl}[FL]{federated learning}
\acrodef{hfl}[HFL]{hierarchical federated learning}
\acrodef{rnn}[RNN]{recurrent neural network}
\acrodef{fcn}[FCN]{fully connected neural network}
\acrodef{lstm}[LSTM]{long short-term memory}
\acrodef{fedavg}[FedAvg]{federated averaging}
\acrodef{ue}[UE]{user equipment}
\acrodef{bs}[BS]{base station}
\acrodef{isp}[ISP]{(wireless) internet service provider}
\acrodef{csp}[CSP]{content service provider}
\acrodef{es}[ES]{edge server}
\acrodef{sgd}[SGD]{stochastic gradient descent}
\acrodef{chr}[CHR]{cache hit ratio}
\acrodef{wgan}[WGAN]{Wasserstein generative adversarial network}
\acrodef{mdp}[MDP]{markov decision process}
\acrodef{csgd}[C-SGD]{centralized SGD}
\acrodef{lru}[LRU]{least recently used}
\acrodef{mru}[MRU]{most recently used}

\begin{document}
\title{Revenue Optimization in Video Caching Networks with Privacy-Preserving Demand Predictions \vspace{-0.1in}}

\author{\IEEEauthorblockN{Yijing Zhang\IEEEauthorrefmark{1}, Md Ferdous Pervej\IEEEauthorrefmark{1} and Andreas F. Molisch}\\
\IEEEauthorblockA{Ming Hsieh Department of ECE, 
University of Southern California, 
CA $90089$, USA \\
Email: {\tt \{yijingz3, pervej, molisch\}@usc.edu} }
\thanks{\IEEEauthorrefmark{1} Equal contributions.}
\vspace{-0.28in}
}

\maketitle
\begin{abstract}
Performance of video streaming, which accounts for most of the traffic in wireless communication, 
 can be significantly improved by caching popular videos at the wireless edge.
Determining the cache content that optimizes performance (defined via a revenue function) is thus an important task, and prediction of the future demands based on past history can make this process much more efficient. 
However, since practical video caching networks involve various parties (e.g., users, \ac{isp}, and \ac{csp}) that do not wish to reveal information such as past history to each other, privacy-preserving solutions are required.
Motivated by this, we propose a proactive caching method based on users' privacy-preserving multi-slot future demand predictions---obtained from a trained Transformer---to optimize \emph{revenue}.
Specifically, we first use a privacy-preserving \ac{fl} algorithm to train a Transformer to predict multi-slot future demands of the users.
However, prediction accuracy is not perfect and decreases the farther into the future the prediction is done.
We model the impact of prediction errors invoking the file popularities, based on which we formulate a long-term system revenue optimization to make the cache placement decisions. 
As the formulated problem is NP-hard, we 
use a greedy algorithm to efficiently obtain an approximate solution. 
Simulation results validate that (i) the \ac{fl} solution achieves results close to the centralized (non-privacy-preserving) solution and (ii) optimization of revenue may provide different solutions than the classical \ac{chr} criterion.
\end{abstract}

\begin{IEEEkeywords}
Cache placement, federated learning, prediction uncertainty, revenue optimization, Transformer.
\end{IEEEkeywords}

%
\IEEEpeerreviewmaketitle

\vspace{-0.1in}

\acresetall
\section{Introduction}


\noindent
Over $70\%$ of total traffic in wireless networks is caused by video streaming \cite{PopcachingSurvey}, the bulk of which is due to repeated requests for a few popular files.
Content caching at the wireless edge, which stores popular content close to the end users (or on the \ac{ue}), can significantly reduce network congestion \cite{golrezaei2013femtocaching}. 
However, limited storage capacity and time-varying content popularity are two key challenges in deciding which files to prefetch into the cache.

While many existing caching algorithms are based on the global popularity (estimated, or assumed known), taking the variations between cells --- regional popularity depends on the preferences of the users within the cell --- into account can significantly improve system performance.  
Besides, existing classical methods of estimating global popularity may not sufficiently capture the dynamic changes in user content requests \cite{pervej2024resource}.
As such, the focus has shifted to cache placement  based on \ac{ml} predictions of the future demands of the individual users in a cell for a particular time window.
While simple \ac{ml} models, such as \ac{rnn}, \ac{lstm}, \ac{fcn}, etc., to name a few, are widely used for these demand predictions (see, e.g., \cite{FedMobileEdgeSurvey} and references therein), recent studies have begun exploring the use of foundation models like Transformers \cite{attentionAllyouNeed} for dynamic caching, motivated by their remarkable performance across various other domains. 

Video streaming involves three parties, i.e., (a) the users who make the requests as encrypted packets using their serving \acp{isp}  (b) the \ac{isp} who forwards the requests to the \ac{csp} without revealing the exact locations of the users, and (c) the \ac{csp} --- often business competitor of the \acp{isp} --- who will not share content information with the \ac{isp}. 
Thus, a 
privacy-preserving collaborative learning solution is required to make  content caching efficient at the wireless edge; this can be achieved based on \ac{fl} \cite{pervej2024resource}.

Once the prediction results are available, the next challenge is to design the caching policy, i.e., determine which content to store in the cache. 
This decision is governed by a tradeoff between the (i) long-term benefits of storing (or replacing) the content set from the storage (due to reduced cost associated with backhauling requested files) and (ii) the cost of refreshing the cache. 
As such, we need to assess the tradeoffs between benefits and costs to define the objective that helps the network to determine the optimal caching strategy.

\vspace{-0.1in}

\subsection{Literature Survey}
\noindent
\Ac{ml} is widely used in content caching to make time-series content predictions that facilitate efficient cache placement. 
Ref. \cite{narayanan2018deepcache} proposed a \ac{lstm} encoder-decoder model
to predict sequential content popularities, 
which then guides the traditional caching strategies, such as \ac{lru}, \ac{mru}, etc.

Ref. \cite{uncertaintyImpact} 
suggests that proactive caching strategies should consider prediction inaccuracies to optimize content delivery and network performance effectively.
Ref. \cite{unrelGaussian} modeled the popularity distribution prediction uncertainty as additive errors on the request probability that follow a Gaussian distribution, and showed the robustness of their caching method to imperfect popularity prediction.

Most cache placement algorithms aim to maximize the \ac{chr}, i.e., the probability that the requested file can be found in the cache; however, network operators are typically more interested in the overall revenue optimization (cost minimization). 
This has been considered in several papers in the past, e.g.,  \cite{Multicell-Coordinated, jointassortpaper, CostMinFuyaru}. 
Ref. \cite{Multicell-Coordinated} formulates the collaborative caching problem to minimize the total cost paid by the content providers to the network operator. 
Similarly, \cite{jointassortpaper} further extends the revenue maximization problem by collaboratively designing personalized assortment decisions and cache planning for wireless content caching networks. 
Ref. \cite{CostMinFuyaru} considers a cost minimization perspective, wherein the system cost contains cache, retrieval, and update costs. Network cache planning was optimized to lower these costs. 

However, all of the above studies have limitations that lead to suboptimal results and/or make them inapplicable for the scenario considered in this paper. 
The predicted results in \cite{narayanan2018deepcache} are still used as input for traditional caching methods, which limits the performance. 
Besides, \cite{narayanan2018deepcache} emphasized predicting content {\em popularity} for each user, i.e., the probability of each video being requested over a long period. 
As a matter of fact, much of the video caching literature is based on (long-term) content {\em popularity}, which may differ from the likelihood of a content being requested in a particular (set of) future request slots. 
In \cite{unrelGaussian}, the authors only show the robustness of their caching algorithm to imperfectly predicted data but do not fully address the unreliable prediction issue.
For the revenue optimization and cost minimization aspect, \cite{Multicell-Coordinated} focuses on file delivery between different \acp{bs},
\cite{jointassortpaper} does not consider revenue optimization in a slot-wise perspective, and \cite{CostMinFuyaru} is based on a given user preference distribution, which is usually unknown in real applications.
Besides, content request popularity in \cite{jointassortpaper} was based on a model in which each user chooses a content that maximizes its own utility. 
Moreover, none of the above works consider privacy preservation aspects.


\vspace{-0.1in}

\subsection{Our contributions}
\noindent
Motivated by these facts, we propose a two-stage proactive caching solution: the users' future requests are predicted for multiple slots using a trained Transformer model, which are then leveraged to optimize revenue-driven cache placement decisions.
Our main contributions are summarized as follows:
\begin{itemize}[leftmargin=*]
\item We use a privacy-preserving \ac{fl} solution \cite{pervej2024resource} that enables collaboration among \ac{csp}, \ac{isp}, and users to predict users' content requests for multiple future slots. Prediction is based on the Transformer structure due to its strong capability in capturing long-term temporal dependencies.
\item We formulate a revenue optimization method that takes the benefits from delivering the requested content to the users and costs for (a) content placement, (b) delivery from \ac{es}' cache, and (c) extraction from the \ac{csp}'s cloud server into account. 
Moreover, since the prediction results from the trained Transformer are not always completely accurate and, hence, cannot directly replace the actual future requests in the revenue objective function, we estimate the actual requests by considering both prediction inaccuracies and user-specific content popularities.
\item Our simulation results demonstrate that the proposed solution has superior performance---in terms of \ac{chr} and revenue---compared to other approaches.
\end{itemize}

\vspace{-0.1in}

\section{System Model}

\begin{figure}
\centering
\includegraphics[trim=0 0 0 5, clip, scale = 0.3]{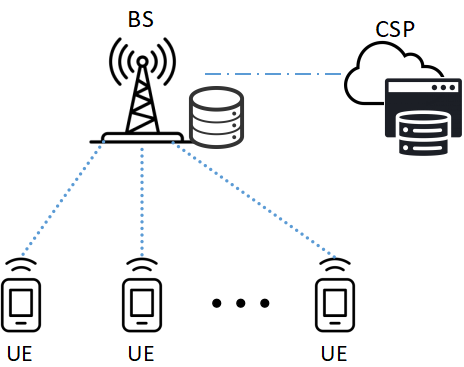}
\caption{Cache placement system model}
\label{fig_systemodel}
\end{figure}
\begin{figure}
\centering
\includegraphics[trim=2 8 35 5, clip, scale = 0.38]{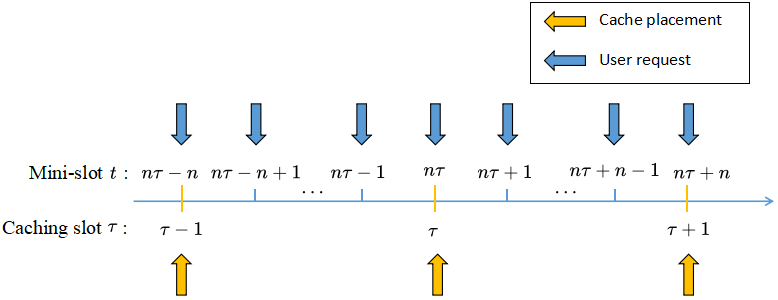}
\caption{User request and cache placement timeslots}
\label{cachingReqSlots}
\end{figure}

\noindent
We consider a wireless video caching network consisting of one \ac{bs}, multiple \acp{ue}, and one \ac{csp}, as illustrated in Fig. \ref{fig_systemodel}.
Let us denote the set of \acp{ue} and content by $ \mathcal{U} = \{u\}_{u=1}^U $ and $ \mathcal{F} = \{f\}_{f=1}^F $, respectively.
We assume that the \ac{bs} can serve all users equally and effectively within its coverage area. 
Each file has the same size, $B$ bits. 
Besides, we use the privacy-preserving collaboration concept of \cite{pervej2024resource} that allows the \ac{csp} to place an \ac{es} at the \ac{bs}.
Denote the storage size of the \ac{es} by $S$, where $S < (F \cdot B)$. 
Furthermore, we assume that content can be placed in \ac{es}' cache only at certain intervals, namely the beginning of what we term a {\em cache placement} slot denoted by $\tau$.
The users, however, can {\em request} content from the \ac{csp} in every $t$ mini-slot, and there are $n$ such mini-slots within each cache placement slot, see Fig. \ref{cachingReqSlots}.
Note that this strategy is widely used \cite{pervej2024efficient} to mimic practical wireless edge caching networks; the specific cache update frequency typically depends on the total amount of traffic in the network and the  application \cite{golrezaei2013femtocaching}.

Denote the content request by a binary variable $i_{u,f}^t\in\{0,1\}$ that takes the value $1$ if user $u$ requests content $f$ during mini-slot $t$ and $0$ otherwise. 
We use a shorthand vector notation to represent all file request information as 
$\mathbf{I}_{u}^t = \{i_{u,1}^t, i_{u,2}^t, \dots, i_{u,F}^t\} \in \mathbb{R}^{F}$.
This work considers that a user only requests a single video content in a given mini-slot. 
As such, we have $\sum_{ f \in \mathcal{F} } i_{u,f}^t=1$, $\forall u \in \mathcal{U}$.
Besides, we denote the cache placement decision by the binary variable $ d_f^\tau\in\{0,1\} $ that takes the value $1$ if the \ac{es} stores content $f$ during cache placement slot $\tau$ and $0$ otherwise.
Furthermore, we assume that placing content in the \ac{es}' cache incurs additional cost, denoted by $c_{\mathrm{plc}}$.
Note that if a requested file is in the \ac{es}' cache, the \ac{bs} and \ac{es} can collaborate based on {\em service-level agreements} to deliver the content to the user locally \cite{pervej2024resource}, which is defined as a {\em cache hit}.
We denote the cost for this local content delivery by $c_{\mathrm{bs-ue}}$.
However, if the requested content is not in the \ac{es}' cache, the \ac{csp} has to extract the content from its cloud server, which is denoted as a {\em cache miss event}.
During a cache miss event, the cost of delivering the requested content to the user is denoted by $c_{\mathrm{bs-ue}} + c_{\mathrm{cl-bs}}$, where $c_{\mathrm{cl-bs}}$ is the cost of extracting the content from the cloud server of the \ac{csp}. 
While both cache placement and request-based retrieval from the cloud require the transmission of the video file on the backhaul link, $c_{\mathrm{cl-bs}}$ is typically larger than $c_{\mathrm{plc}}$ because cache placement is typically done at times of lower general traffic load and thus lower cost for using up bandwidth. 
Finally, the \emph{benefit} of delivering the requested content to the requesting user is $\beta$, regardless of a cache \emph{hit} or \emph{miss} event.

\vspace{-0.1in}
\section{Revenue Optimization Cache Planning based on Transformer Prediction}
\noindent
Intuitively, storing the \emph{to-be-requested} content set for a long time is beneficial since caching a video file has the cost $c_{\mathrm{plc}}$, which is not only typically lower than $c_{\mathrm{cl-bs}}$, but also non-recurring (i.e., repeated requests for a cached file do not result in additional costs for the actual caching). 
However, the (global or regional) content popularity can change over a short period (even if the overall popularity distribution is constant over a long time horizon):
the probability of cache miss events can increase with a static cache, i.e., the cost associated with backhaul extractions can be significantly higher. 
Since the number of files in a cache is limited, this motivates cache refreshes. 
As such, it is critical to maximize system revenue from a long-term perspective. 
Note that these costs and benefits can be interpreted in terms of money from the network operator's perspective. The costs represent the expenses incurred by the operator for renting backhaul and over-the-air transmission resources\footnote{If the network operator {\em owns} the transmission resources, it can be seen as the fraction of the depreciation over time.} necessary for the transmission of the video, while the benefit corresponds to the fee paid by user and/or \ac{csp} to the \ac{isp} for this delivery.

\vspace{-0.15in}
\subsection{Cache Planning for Revenue Optimization}

\noindent
In this work, we focus on slot-wise cache planning. 
As such, we first define the revenue of our video caching network at cache placement slot $\tau$ as follows:
\begin{align}
\label{eq:R_total}
&R_{\mathrm{total}}^{\tau} 
\coloneqq R_{\mathrm{req}}^{\tau} + R_{\mathrm{plc}}^{\tau} + \gamma R_{\mathrm{plc}}^{\tau+1}, \nonumber\\
&\quad = \underbrace{\sum\nolimits_{t = n\tau}^{ n(\tau+1)-1}\sum\nolimits_{f=1}^F \sum\nolimits_{u=1}^U i_{u,f}^t( \beta - c_{\mathrm{bs-ue}} - (1-d_f^{\tau}) c_{\mathrm{cl-bs}})}_{R_{\mathrm{req}}^\tau} \nonumber\\
&\quad \underbrace{- c_{\mathrm{plc}} \sum\nolimits_{f=1}^F d_f^{\tau} (1-d_f^{\tau-1})}_{R_{\mathrm{plc}}^{\tau}} + 
\gamma \cdot 
\underbrace{c_{\mathrm{plc}} \sum\nolimits_{f=1}^F d_f^{\tau+1} d_f^{\tau}}_{R_{\mathrm{plc}}^{\tau+1}}, 
\end{align}
where the first term is the cumulative net benefit of delivering the requested content to the users, taking into account transmission and backhaul costs.
The second term is the cost of changing the content in the \ac{es}' cache for the current slot, while the last term is the benefit of having the files in the cache for future timeslots, where $\gamma\in(0,1)$ denotes the future discount factor. 
While one might include discounted future revenues $R_{\mathrm{req}}^{\tau}$, we approximate them as independent of the cache configuration at time $\tau$ (see remark 1), so that they do not impact optimization of $d_f^{\tau}$.  

We now focus on finding cache placement decisions, $\{d_f^\tau\}_{f=1}^F$, so that the network maximizes its revenue. 
As such, we pose our optimization problem as follows:
\begin{subequations}
\label{optproblem}
\begin{align}
\underset{\{d_f^{\tau}\}_{f=1}^F} {\mathrm{maximize}} & \qquad  R_{\mathrm{total}}^{\tau} \tag{\ref{optproblem}} \\
\mathrm{subject ~to} &\qquad C1:\sum\nolimits_{f=1}^F d_f^{\tau} B \leq S, \\ 
&\qquad C2: d_f^{\tau}\in\{0,1\}, \quad \forall f \in \mathcal{F},
\end{align}
\end{subequations}
where constraints $C1$ and $C2$ are due to the limited cache storage size of the \ac{es} and binary cache placement decision variables, respectively.
The optimization problem in (\ref{optproblem}) is a $0-1$ Knapsack problem, which is NP-hard.

\begin{remark}
  Eq. (\ref{eq:R_total}) is not an exact formulation. Rather, one should instead define $R_{\mathrm{total}}^{\tau}=\sum_{\tau'=\tau}^{\infty} \gamma_{\rm ts}^{\tau'-\tau} \left(R_{\mathrm{req}}^{\tau} + R_{\mathrm{plc}}^{\tau} \right )$. However, this problem is highly challenging to solve directly because the cache has to be updated at the cache placement slot $\tau$, while the revenue depends on the future content requests $i_{u,f}^t$ in all mini-slots $t \in [n\tau, n(\tau+1)-1]$, which are unknown.
Furthermore, the cache placement at $\tau+1$ depends on all future requests, so that in principle one would have to consider longer cache placement slots prediction and solve this problem as a \ac{mdp} or multi-stage Knapsack problem. We will tackle this problem in our future work. 
\end{remark}

To solve the challenge of knowing the future requests (including their probabilities) in the mini-slots of the current cache placement timeslot, in the sequel, we will use a privacy-preserving \ac{fl} algorithm to train a Transformer \cite{attentionAllyouNeed} that provides the demand prediction combined with an approximate probabilistic model for the actual requests when the transformer prediction fails. 
To deal with the unknown requests in future cache placement timeslots $[\tau+1,\tau+2,....]$, we assume that future cache placement decisions $\{d_f^{\tau+1}\}_{f=1}^F$ are made based on the global content popularity $\mathbf{G}_f$ to approximately optimize $d_f^\tau$ in each cache placement slot $\tau$. 
Considering that the prediction reliability for a full caching slot into the future is typically low, this is a reasonable approximation that drastically simplifies the problem. 

\vspace{-0.1in}
\subsection{Two-Stage Approximate Revenue Optimization}
\noindent
Due to the above complexities with the original problem, we use a two-stage solution. 
First, we use a Transformer \cite{attentionAllyouNeed} to learn multi-step demand predictions by training it with the \ac{fedavg} \cite{McMahanFL} algorithm. 
Then, the trained Transformer's predictions are used to approximately replace the original unknown objective function. 

\subsubsection{Stage 1: \ac{fl}-aided Demand Predictions}

We leverage \ac{ml} to predict users' content requests for all $t \in [n\tau, n(\tau+1)-1]$ mini-slots, which will be used to make cache placement decisions at the beginning of the cache placement slot $\tau$.
In particular, we want to train a Transformer, parameterized by weight vector $\Theta$.
However, since data only belong to the users and cannot be shared due to privacy concerns, we use the \ac{fedavg} \cite{McMahanFL} algorithm to train our model $\Theta$ using the following key steps.

At the beginning of each (global) training round $r$, the \ac{es} broadcasts the global model $\Theta^r$ to all $\mathcal{U}$. 
The users then synchronize their local model as $\Theta_u^0 \gets \Theta^r$ and take $K$ mini-batch \ac{sgd} steps as $\Theta_u^{K} = \Theta_u^0 - \eta \sum_{\tau=0}^{K-1} \nabla l_u \left(\Theta_u^\tau | (\zeta \sim \mathcal{D}_u) \right)$, where $\eta$ is the learning rate and $l_u$ is the loss function. 
All users then offload their updated local model to the \ac{es}, which then aggregates these models as $\Theta^{r+1} = \frac{1}{U}\sum_{u=1}^U \Theta_u^K$. 
The \ac{es} then broadcasts this updated model and the users again follow the above steps. 
This process is repeated for $r_g$ global rounds.

We assume that the above training happens offline, and during the online test phase, the users use the trained global model $\Theta^{*}$ to predict their future requests. As we want to predict content requests in the next $n$ mini-slots that start at $t$, the input feature is $\mathbf{x}_u = (\mathbf{I}_{u}^{t-N}, \mathbf{I}_{u}^{t-N+1}, \cdots, \mathbf{I}_{u}^{t-1}) \in \mathbb{R}^{N\times F}$ from prior $N$ mini-slot content requests, the output for next $n$ mini-slot prediction is $\mathbf{\hat{y}}_u = (\mathbf{\hat{I}}_{u}^{t}, \mathbf{\hat{I}}_{u}^{t+1}, \cdots, \mathbf{\hat{I}}_{u}^{t+n-1}) \in \mathbb{R}^{n\times F}$. It is worth noting that $\hat{i}_{u,f}^t \in [0,1]$ from $\mathbf{\hat{I}}_{u}^{t}$ is essentially the {\em probability} that the user requests content $f$ during mini-slot $t$.
Then the information predicted at the \ac{ue} will be calculated (including addressing the uncertainty as mentioned in the next section), and sent to the \ac{es} as input for the revenue optimization.
\noindent

\subsubsection{Stage 2: Transformer's Prediction Assisted Revenue Optimization}
While the straightforward way to approximately solve (\ref{optproblem}) is to use predictions $\hat{i}_{u,f}^t$ to replace the objective function directly, such a naive approach neglects the impact of the prediction accuracy.
Since it is reasonable to consider that the prediction results from any \ac{ml} model can be imperfect, it is intuitive to balance the unreliable predictions of our trained $\Theta^{*}$. 
As such, we use the prediction results $\hat{i}_{u,f}^t$ from the trained $\Theta^{*}$, {\em weighted by the prediction accuracy} $a_{u,f}^t$ to approximate the content request for solving (\ref{optproblem}).

To that end, for each \emph{mini-slot} $t$ within \emph{cache placement slot} $\tau$, we calculate the prediction accuracy $a_{u,f}^t$ using the validation dataset and prediction results of the users as 
\begin{equation*}
    a_{u,f}^{t} \coloneqq \frac { \sum \nolimits_{\tau'=1}^{K} \delta \left( f - \mathrm{argmax} [\hat{\mathbf{I}}_{u}^t] \right) \times \delta \left( f - \mathrm{argmax} [\mathbf{I}_{u}^t] \right) } {\sum_{\tau'=1}^{K} \delta \left( f - \mathrm{argmax} [\mathbf{I}_{u}^t] \right)}, \forall t \in \tau'
\end{equation*}
where $K$ is the total \emph{cache placement slots} in the validation set, and $\delta(\cdot)$ is a delta function. 
Note that $a_{u,f}^t$, i.e., the accuracy of a particular file being requested at a specific mini-slot $t$, is essentially calculated by taking the fraction of the events that $\Theta^*$ correctly predicted file $f$ is requested in mini-slot $t$ over the total number of events that indeed file $f$ is requested in that specific mini-slot $t$ in all $K$ cache placement slots. 
Denote the popularity of the $f^{\mathrm{th}}$ content at user $u$ by $g_{u,f}$ such that $\sum_{f=1}^F g_{u,f}=1$: users can calculate this popularity using their historical information.

Then, we strike a balance between the Transformer's prediction accuracy and \emph{user-specific} content popularity to estimate the actual content request $i_{u,f}^t$ as follows.
\begin{proposition}
\label{prop}
Based on the prediction accuracy $a_{u,f}^t$ and predicted request $\hat{i}_{u,f}^t$ from the transformer, the estimation of actual content request is written as
\begin{equation}
\label{eq:estimation_true}
i^{t}_{u,f,\mathrm{est}} = \hat{i}_{u,f}^t a_{u,f}^t + g_{u,f} (1-a_{u,f}^t),
\end{equation}
where $\hat{i}_{u,f}^t$ is the $f^{\mathrm{th}}$ entry of the vector $\hat{\mathbf{I}}_u^t$.
\end{proposition}


Based on the estimation of the actual content request, i.e., $i_{u,f}^t \approx i_{u,f,\mathrm{est}}^t$, we now rewrite the approximate revenue as
\begin{align}
\label{eq:revObjOur}
&R_{\mathrm{total}}^{\tau} \approx \sum\nolimits_{t = n\tau}^{n(\tau+1)-1}\sum\nolimits_{f=1}^F \sum\nolimits_{u=1}^U \big[ i^{t}_{u,f,\mathrm{est}} \cdot ( \beta - c_{\mathrm{bs-ue}} - c_{\mathrm{cl-bs}}) \big] \nonumber
\\& + \sum\nolimits_{t = n\tau}^{ n(\tau+1)-1} \sum\nolimits_{f=1}^F \sum\nolimits_{u=1}^U i^{t}_{u,f,\mathrm{est}} \cdot d_f^{\tau} c_{\mathrm{cl-bs}} -c_{\mathrm{plc}}  \sum\nolimits_{f=1}^F d_f^{\tau} 
+ \nonumber\\
&\quad c_{\mathrm{plc}}  \sum\nolimits_{f=1}^F d_f^{\tau} d_f^{\tau-1}
+\gamma \cdot c_{\mathrm{plc}}  \sum\nolimits_{f=1}^F d_f^{\tau+1} d_f^{\tau} \coloneqq \tilde{R}_{\mathrm{total}}^{\tau}.
\end{align}
Note that (\ref{eq:revObjOur}) only transforms the objective function of (\ref{optproblem}): the problem is still NP-hard.
With slight algebraic manipulations, we rewrite the original problem (\ref{optproblem}) as
\begin{subequations}
\label{optproblem_Transformed}
\begin{align}
\underset{\{d_f^{\tau}\}_{f=1}^F} {\mathrm{maximize}} & \qquad  \tilde{R}_{\mathrm{total}}^{\tau} = \sum\nolimits_{f=1}^{F} d_f^{\tau} W_f + Z, \tag{\ref{optproblem_Transformed}} \\
\mathrm{subject ~to} &\qquad C1, ~C2,
\end{align}
\end{subequations}
where 
$W_f = \! \! \sum\nolimits_{t = n\tau}^{ n(\tau+1)-1} \sum\nolimits_{u=1}^U \! i^{t}_{u,f,\mathrm{est}} c_{\mathrm{cl-bs}} 
 - c_{\mathrm{plc}} \! + c_{\mathrm{plc}}  d_f^{\tau-1} 
 \! + \gamma c_{\mathrm{plc}}   d_f^{\tau+1}$
and $Z = \sum\nolimits_{t = n\tau}^{ n(\tau+1)-1}\sum\nolimits_{f=1}^F \sum\nolimits_{u=1}^U i^{t}_{u,f,\mathrm{est}} \cdot( \beta - c_{\mathrm{bs-ue}} - c_{\mathrm{cl-bs}})$.
It is easy to check that $W_f$ and $Z$ are constants. 
When we assume all content are the same size, this problem 
can be solved by using a greedy algorithm \cite[Chapter 2]{knapsackBook}, which chooses the file with the highest value $W_f$ to the lowest.

Finally, given that the caching decision, $\{d_f^\tau\}_{f=1}^F$, $\forall f$, and $\tau$, are to be determined for each cache placement slot sequentially, we now summarize all steps in our proposed two-stage revenue optimization technique 
as follows. 

\textbf{Step 1}: 
Each \ac{ue} uses the global model $\Theta^{*}$ and its own request history  to predict the next $n$ mini slots' content requests probability $\hat{i}_{u,f}^t$.

\textbf{Step 2}: 
Each \ac{ue} calculates the estimated content request ${i}_{u,f,\mathrm{est}}^t$ using (\ref{eq:estimation_true}) and then sends\footnote{It is worth noting that users only share their estimation results, i.e., users do not reveal their true future request information to the ISP.} this information to the \ac{es}.

\textbf{Step 3}: The \ac{es} aggregates all user-estimated requests $i_{u,f,\mathrm{est}}^t$ and forms the revenue optimization problem.

\textbf{Step 4}: 
The \ac{es} then solves the revenue optimization problem in (\ref{optproblem_Transformed}) using \cite[Chapter $2$]{knapsackBook} and get optimized cache decisions $\{d_f^{\tau^*}\}_{f=1}^F$. 


\vspace{-0.1in}
\section{Simulation Results and Discussions} 

\subsection{Simulation Settings}

\noindent
Since no suitable spatio-temporal dataset is publicly available \cite{pervej2024resource}, we use the following content request model for our performance evaluation.
We assume that the files in the catalog are grouped by genre, denoted by $ \mathcal{G} = \{\mathfrak{g}\}_{\mathfrak{g}=1}^G $. 
Each content within the genre $\mathfrak{g}$ has popularity $P_{\mathfrak{g},f}$, which follows a Zipf distribution with exponent $\tilde{\gamma}$, and $\sum \nolimits_{f\in \mathfrak{g}} P_{\mathfrak{g},f}=1$. 
The users have their own genre preferences, denoted by $p_{u,\mathfrak{g}}$, where $\sum \nolimits_{\mathfrak{g}=1}^G p_{u,\mathfrak{g}}=1$.
We assume the user request model follows a day-by-day behavior. For a single day, a user selects the genre $\mathfrak{g}$ with probability $p_{u,\mathfrak{g}}$ and remains within that genre for the entire day. 
For requests generation, each user first randomly requests $L$ distinct content based on content popularity, similarity, and time dependency. 
With slight abuse of notations, suppose these $L$ contents requested by user $u$ are  $\mathcal{F}_{u,L} = \{ f_{1}, f_{2},...,f_{l},...,f_{L}\}$. Denote the feature set of the $f_l^{\mathrm{th}}$ content by $\pmb{\phi}_{f_l}$.
We then compute the similarity score of all other files $f'' \in \mathcal{F} \setminus \mathcal{F}_{u,L}$ to these $L$ files, with recent history contributing more weight, as
\begin{equation*}
    S_{u,f''} = \sum \nolimits_{l = 1}^{L} \exp \left[ -(L-l)/b\right] \cdot \left(\pmb{\phi}_{f_l} \cdot \pmb{\phi}_{f''} \right)/\left(\Vert \pmb{\phi}_{f_l} \Vert \ \Vert \pmb{\phi}_{f''}\Vert \right).
\end{equation*}
where 
$b$ is a constant. 
Then, the subsequent $M$ requests are generated from $\mathcal{F} \setminus \mathcal{F}_{u,L}$ with Top-$M$ probabilities, which are calculated as
\begin{equation*}
    \bar{S}_{u,f''} = \lambda \cdot \frac{\mathrm{exp}(S_{u,f''})}{\sum\limits_{f'' \in \mathcal{F} \setminus \mathcal{F}_{u,L} } \!\!\!\! \mathrm{exp}(S_{u,f''})} + (1-\lambda) \cdot \frac{\mathrm{exp}(P_{\mathfrak{g},f''})}{\sum\limits_{f'' \in \mathcal{F} \setminus \mathcal{F}_{u,L}} \!\! \!\! \mathrm{exp}(P_{\mathfrak{g},f''})}
\end{equation*}
where $\lambda\in (0,1)$ is the weighting factor. 
Each user repeats this procedure until the total number of requests is reached for that day, and follows the same procedure for the next day.

Based on the above request model, we consider a video caching network consisting of $U=50$ users and $F=240$ files\footnote{While our simulation setting is relatively small compared to real-world scenarios (e.g., Netflix, which hosts hundreds of thousands of video contents) due to having reasonable computation time, the revenue optimization principle shall remain the same regardless of the number of content.}; each file has a size $B = 1$. 
Besides, we consider $G=3$, $L = 7$, $M=5$, $b=0.5$. Content popularity in genre $P_{\mathfrak{g},f}$ follows a Zipf distribution with exponent $\tilde{\gamma}=1.5$. User genre preferences $p_{u,\mathfrak{g}}$'s are modeled using a symmetric Dirichlet distribution $\mathrm{Dir}(\alpha_u)$, where $\alpha_u = 0.3$ is the concentration parameter. The total number of requests for each user on a single day is $107$.

Each user uses their historical content request information $\mathbf{I}_{u}^t$, from the past $160$ days, to prepare their respective training dataset $\mathcal{D}_u = \{\mathbf{x}_u^m,\mathbf{y}_u^m\}_{m=1}^{D_u}$, where ${D_u}$ is the total number of training samples. Denoting $t_m$ as the first mini-slot of the $m^{\mathrm{th}}$ training sample label, we have $\mathbf{x}_u^m = (\mathbf{I}_{u}^{t_m-N}, \mathbf{I}_{u}^{t_m-N+1}, \cdots, \mathbf{I}_{u}^{t_m-1}) \in \mathbb{R}^{N\times F}$ and $\mathbf{y}_u^m = (\mathbf{I}_{u}^{t_m}, \mathbf{I}_{u}^{t_m+1}, \cdots, \mathbf{I}_{u}^{t_m+n-1}) \in \mathbb{R}^{n\times F}$ as the feature set (contains $N$ historical content requests) and label set (contains $n$ future requests) of the $m^{\mathrm{th}}$ training sample, respectively.

The Transformer model has $6$ encoder/decoder layers. 
We use embedding
dimension $512$, feed-forward dimension $1024$, with single head attention for predicting $n=2$ slots, and $2$ heads attention for predicting $n=5$ slots.
The input data for the decoder layer are padded with zero vectors, which work as placeholders for the predicted values \cite{jiang2022accurate}.
For \ac{fl} training, each user trains their local model for $4$ epochs before offloading the trained model to the \ac{es}.
The \ac{es} puts equal weights during model aggregations and trains the model for $450$ global rounds. 
Besides, we use cross-entropy loss function and SGD optimizer with learning rate $0.18$. 
Furthermore, our revenue optimization parameter settings are $\beta= 3$, $c_{\mathrm{bs-ue}} = 0.5$, $c_{\mathrm{cl-bs}} = 2$, $\gamma = 0.8$; $c_{\mathrm{plc}} = 0.7$ for $n=2$ and $c_{\mathrm{plc}} = 1$ for $n=5$. We consider cache size $S$ ranging from $20$ to $240$ to examine the effects of different cache sizes.

\begin{table}[t!]\small
\label{table:FLCL}
\centering
\caption{Performance comparison between CL and FL}
\begin{tabular}{ |c|c| } 
 \hline
    \bf Method & \textbf{Test Accuracy}  \\ 
    \hline
 Centralized Learning & $0.8576 \pm 0.0073$  \\ 
 \hline
 Federated Learning & $0.8531 \pm 0.0045$  \\ 
 \hline
\end{tabular}
\label{tab:fl_vs_sgd}
\end{table}
\begin{figure*}[!ht]  
  \centering
  \begin{subfigure}[t]{0.24\textwidth}
    \centering
    \includegraphics[scale = 0.3]{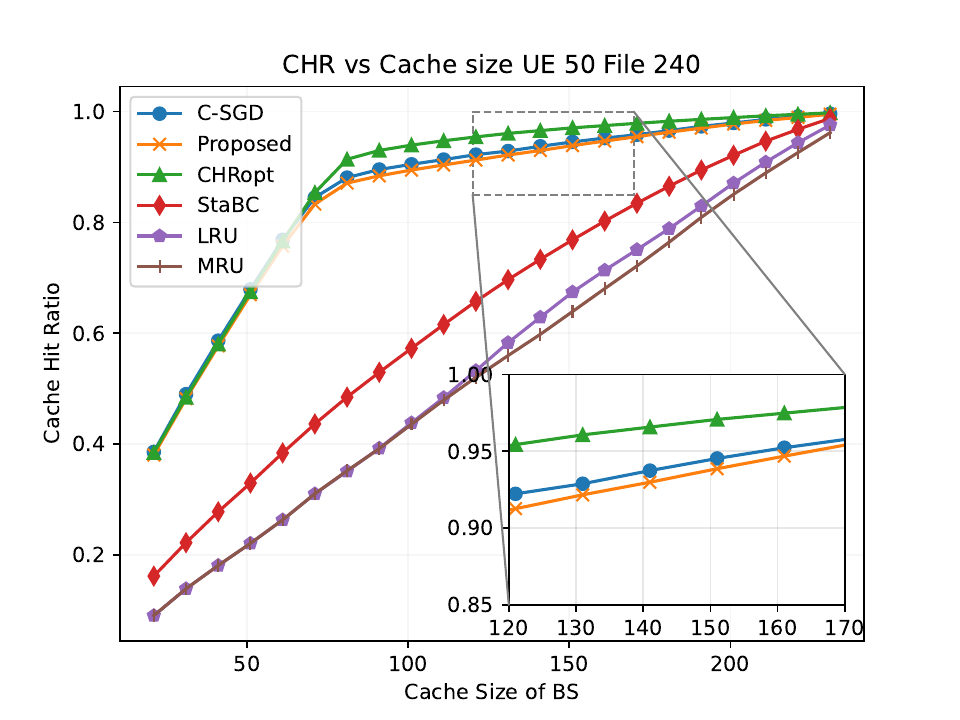}
    \caption{CHR comparisons: $n = 2$}
    \label{fig:chr_n2}
  \end{subfigure}\hfill%
  \begin{subfigure}[t]{0.24\textwidth}
    \centering
    \includegraphics[scale = 0.3]{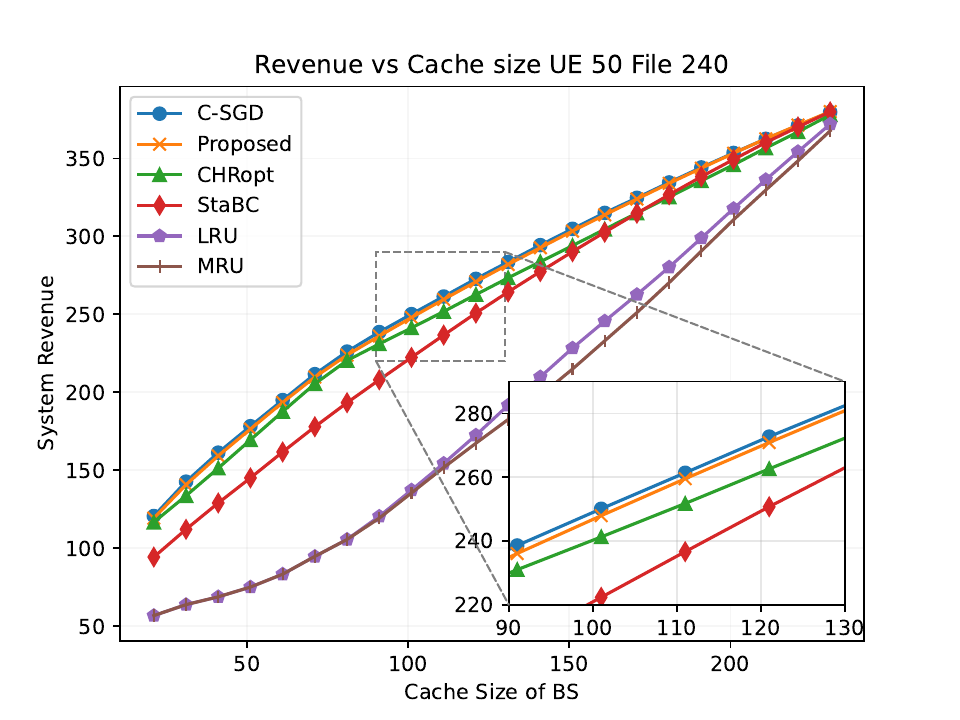}
    \caption{Revenue comparisons: $n = 2$}
    \label{fig:rev_n2}
  \end{subfigure}\hfill%
  \begin{subfigure}[t]{0.24\textwidth}
    \centering
    \includegraphics[scale = 0.3]{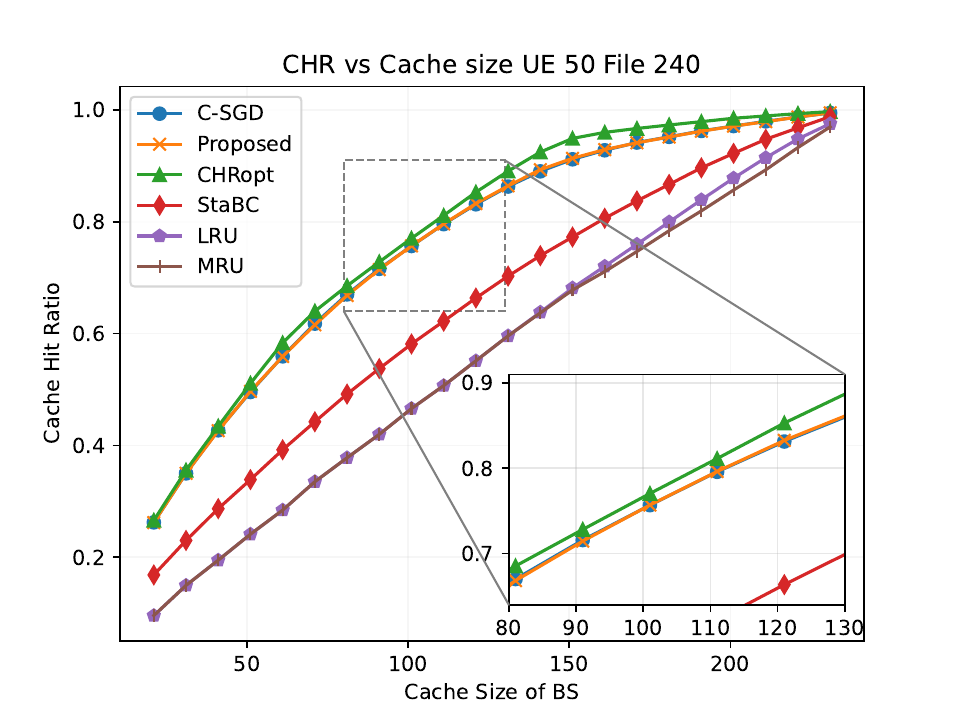}
    \caption{CHR comparisons: $n = 5$}
    \label{fig:chr_n5}
  \end{subfigure}\hfill%
  \begin{subfigure}[t]{0.24\textwidth}
    \centering
    \includegraphics[scale = 0.3]{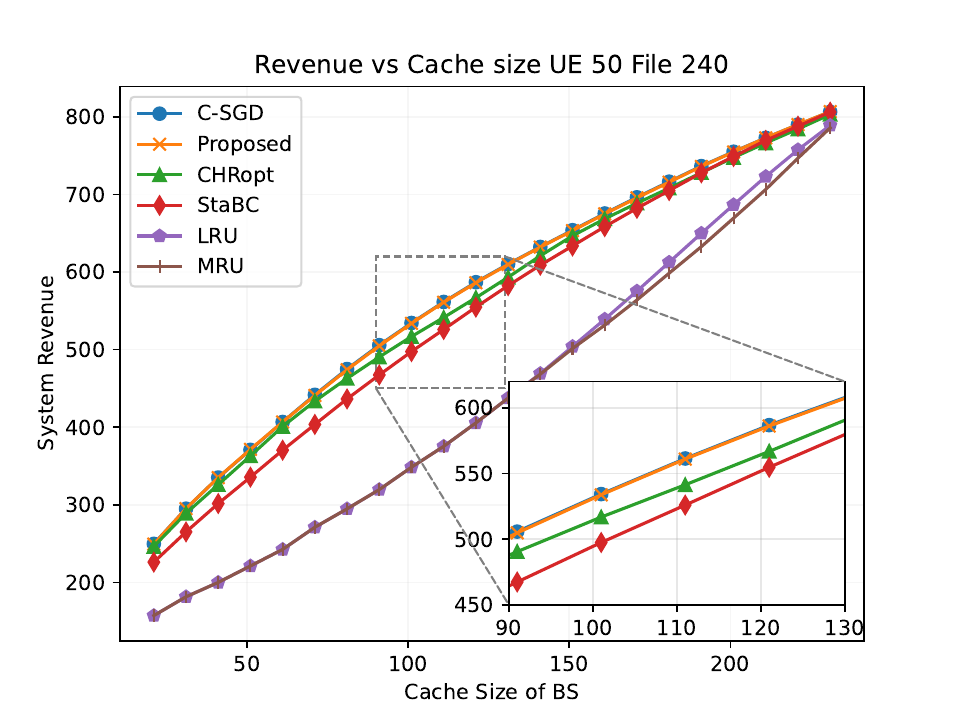}
    \caption{Revenue comparisons: $n = 5$}
    \label{fig:rev_n5}
  \end{subfigure}
  \caption{Cache hit ratio and revenue comparison in $n=2$ and $n=5$ for different cache placement methods}
  \label{fig:chr_revenue_n_5}
  \vspace{-0.25in}
\end{figure*}

\vspace{-0.15in}
\subsection{Performance Comparisons}
\noindent
At first, we check the performance gap between \ac{fl} and a \ac{csgd}, assuming a \emph{Genie} knows all training datasets from all users, method. 
Since the Genie has access to all datasets, we consider its performance to indicate the upper bound of \ac{ml} methods.  
As such, when we have privacy concerns and are required to use \ac{fl} to train our model, we expect the globally trained model to suffer from statistical data heterogeneity.
Our results in Table \ref{tab:fl_vs_sgd} also validate this: \ac{fedavg} delivers $99.4\%$ of the \ac{csgd}'s test accuracy.

We then investigate how the prediction accuracy behaves across different mini-slots in our video caching task. 
As the content request models follow a popularity-preference tradeoff, predicting the exact requests in all future mini-slots can be challenging. 
Our simulation results show that the accuracies are $0.8531 \pm 0.0045$, $0.8167 \pm 0.0046$, $0.7968 \pm 0.0051$, $0.7771 \pm 0.0061$, and $0.7414 \pm 0.0047$, in mini-slots $1$ to $5$, respectively, yielding an average test accuracy of $0.7970 \pm 0.0049$. 
Given these facts and results, we now focus on performance comparisons with other baselines.

\subsubsection{Baselines}
\noindent
We consider the following \emph{cache placement} baselines to show the effectiveness of our proposed two-stage cache placement solution that maximizes the system revenue. 

\noindent
\textbf{Cache Hit Ratio Criteria (\textbf{CHRopt})}: This baseline stores files that maximize $\sum\nolimits_{t = n\tau}^{n(\tau+1)-1}\sum\nolimits_{u=1}^U i^{t}_{u,f,\mathrm{est}}$ at each cache placement slot.

\noindent 
\textbf{Statistics-based cache placement (\textbf{StaBC})}: Content stored at \ac{es}' cache based on historical global content popularity.

\noindent 
\textbf{Least recently used (\textbf{LRU})}: The \ac{es} tracks all users' historical requests up to cache placement slot $\tau-1$. Requested content during $[n\tau-n, n\tau-1]$ but absent from the cache will be stored, replacing the least recently used content\footnote{Note the definition of \ac{lru} and \ac{mru} may be different from other work.}.

\noindent
\textbf{Most recently used (\textbf{MRU})}: The \ac{es} tracks all users' historical requests up to cache placement slot $\tau-1$. Requested content during $[n\tau-n, n\tau-1]$ but absent from the cache will be stored, replacing the most recently used content.


\subsubsection{Performance Analysis}

We use the average (a) \ac{chr} and (b) revenue per cache placement slot as evaluation criteria.
We consider two cases: (a) $n=2$ and (b) $n=5$, and vary the cache size $S$ to investigate the performances.

Intuitively, having a larger cache size means more content can be placed in the cache. 
As such, as $S$ increases, both \ac{chr} and revenue are expected to increase for all cache placement methods, which is trivial.
However, given a fixed $S$, we expect (\ref{optproblem_Transformed}) to yield optimized cache placement decisions considering the tradeoff between the likelihood of a file being used according to the Transformer prediction, and it being used because it has a higher popularity, due to our estimation method in (\ref{eq:estimation_true}). In line with our expectation, we find our optimized (non-privacy-preserving) C-SGD to perform best, with the (privacy-preserving) FL-based approach providing almost identical results. Since our proposed method focuses on optimizing long-term revenue, we expect that it may have lower performance for the \ac{chr} but the best performance for the achievable revenue.
The CHRopt baseline, on the other hand, decides cache placement solely by maximizing the probability that future requested content hits the cache. As a result, although it may perform well in terms of \ac{chr}, it achieves a lower revenue compared to the proposed method.
The naive baselines (StaBC, LRU, and MRU) are expected to perform worse. 
This is due to the fact that the statistical cache placement only considers global popularities without considering individual user preferences.
As instantaneous file requests evolve over time, the popularity-based caching approach fails to adapt to these changes.
Moreover, since it relies solely on historical data, it is inherently less effective than our prediction-based method, which more accurately anticipates future demands.
Similarly, LRU and MRU only emphasize the previous content requests, while future content requests might differ.

Our simulation results in Fig. \ref{fig:chr_revenue_n_5} also show the above trends.
For both $n=2$ and $n=5$, we observe that our proposed solution significantly outperforms the other baseline in terms of both CHR and system revenue.
For example, when $n=5$, $S=100$, the \acp{chr} are about $75\%$, $77\%$, $58\%$, $46\%$, and $46\%$, with our proposed solution, CHRopt, StaBC, LRU, and MRU, respectively. 
Besides, the corresponding system revenues are about $534$, $516$, $497$, $348$, and $348$, respectively.



\vspace{-0.1in}
\section{Conclusions} 
\noindent
We proposed a privacy-preserving cache placement method for wireless video caching networks using FL-trained Transformers for multi-step content prediction. The predictions, weighted by accuracy and global popularity, optimized cache placement to maximize system revenue. Our  solution outperformed baselines in CHR and revenue. Future work will focus on enhancing revenue optimization by considering long-term caching strategies and stochastic request uncertainties.

\vspace{-0.1in}
\bibliographystyle{ieeetr} 
\bibliography{refs} 

\begin{thebibliography}{10}

\bibitem{PopcachingSurvey}
H.~S. Goian {\em et~al.}, ``Popularity-based video caching techniques for cache-enabled networks: A survey,'' {\em IEEE Access}, vol.~7, pp.~27699--27719, 2019.

\bibitem{golrezaei2013femtocaching}
N.~Golrezaei, A.~F. Molisch, A.~G. Dimakis, and G.~Caire, ``Femtocaching and device-to-device collaboration: A new architecture for wireless video distribution,'' {\em IEEE Commun. Magaz.}, vol.~51, no.~4, pp.~142--149, 2013.

\bibitem{pervej2024resource}
M.~F. Pervej and A.~F. Molisch, ``Resource-aware hierarchical federated learning in wireless video caching networks,'' {\em IEEE Trans. Wireless Commun.}, vol.~24, no.~1, pp.~165--180, 2025.

\bibitem{FedMobileEdgeSurvey}
W.~Y.~B. Lim {\em et~al.}, ``Federated learning in mobile edge networks: A comprehensive survey,'' {\em IEEE Commun. Surveys Tutor.}, vol.~22, no.~3, pp.~2031--2063, 2020.

\bibitem{attentionAllyouNeed}
A.~Vaswani {\em et~al.}, ``Attention is all you need,'' in {\em Proc. Advances NeurIPS}, 2017.

\bibitem{narayanan2018deepcache}
A.~Narayanan, S.~Verma, E.~Ramadan, P.~Babaie, and Z.-L. Zhang, ``Deepcache: A deep learning based framework for content caching,'' in {\em Proc. 2018 Workshop on Network Meets AI \& ML}, pp.~48--53, 2018.

\bibitem{uncertaintyImpact}
P.~Cong, K.~Qi, and C.~Yang, ``Impact of prediction uncertainty of popularity distribution on proactive caching,'' in {\em Proc. IEEE/CIC ICCC}, 2019.

\bibitem{unrelGaussian}
H.~Kim, J.~Park, M.~Bennis, S.-L. Kim, and M.~Debbah, ``Mean-field game theoretic edge caching in ultra-dense networks,'' {\em IEEE Trans. Vehicular Technol.}, vol.~69, no.~1, pp.~935--947, 2020.

\bibitem{Multicell-Coordinated}
A.~Gharaibeh, A.~Khreishah, B.~Ji, and M.~Ayyash, ``A provably efficient online collaborative caching algorithm for multicell-coordinated systems,'' {\em IEEE Trans. Mobile Comput.}, vol.~15, no.~8, pp.~1863--1876, 2016.

\bibitem{jointassortpaper}
Y.~Fu {\em et~al.}, ``Joint assortment and cache planning for practical user choice model in wireless content caching networks,'' {\em IEEE Trans. Mobile Comput.}, vol.~23, no.~5, pp.~4709--4722, 2024.

\bibitem{CostMinFuyaru}
Y.~Fu {\em et~al.}, ``Optimal and suboptimal dynamic cache update algorithms for wireless cellular networks,'' {\em IEEE Wireless Commun. Letters}, vol.~11, no.~12, pp.~2610--2614, 2022.

\bibitem{pervej2024efficient}
M.~F. Pervej, R.~Jin, S.-C. Lin, and H.~Dai, ``Efficient content delivery in user-centric and cache-enabled vehicular edge networks with deadline-constrained heterogeneous demands,'' {\em IEEE Trans. on Vehicular Technol.}, vol.~73, no.~1, pp.~1129--1145, 2024.

\bibitem{McMahanFL}
B.~McMahan, E.~Moore, D.~Ramage, S.~Hampson, and B.~A.~y. Arcas, ``{Communication-Efficient Learning of Deep Networks from Decentralized Data},'' in {\em Proc. AIStat}, 2017.

\bibitem{knapsackBook}
S.~Martello and P.~Toth, {\em Knapsack problems: algorithms and computer implementations}.
\newblock USA: John Wiley \& Sons, Inc., 1990.

\bibitem{jiang2022accurate}
H.~Jiang, M.~Cui, D.~W.~K. Ng, and L.~Dai, ``Accurate channel prediction based on transformer: Making mobility negligible,'' {\em IEEE J. Sel. Areas Commun.}, vol.~40, no.~9, pp.~2717--2732, 2022.

\end{thebibliography}

\end{document}